\documentclass{article}
\usepackage{tex-style/spconf}
\usepackage{cite}
\usepackage{amsmath,amssymb,amsfonts}
\usepackage{algorithmic}
\usepackage{graphicx}
\usepackage{textcomp}
\usepackage{xcolor}
\usepackage{amsmath}
\usepackage[]{algorithm2e}
\usepackage{diagbox}
\usepackage{todonotes}
\usepackage{blindtext}
\usepackage{caption}
\SetAlCapSkip{1em}

\title{Encoder complexity control in SVT-AV1 by speed-adaptive preset switching}
\name{
    \parbox{\linewidth}{\centering
    Lena Eichermüller$^{\star}$ 
    \qquad Gaurang Chaudhari$^{\dagger}$
    \qquad Ioannis Katsavounidis$^{\dagger}$  
    \qquad Zhijun Lei$^{\dagger}$ \quad
    \qquad Hassene Tmar$^{\dagger}$
    \qquad André Kaup$^{\star}$ 
    \qquad Christian Herglotz$^{\star}$}}

\address{$^{\star}$ Multimedia Communications and Signal Processing \\ 
    Friedrich-Alexander-Universität Erlangen-Nürnberg, Erlangen, Germany \\
    $^{\dagger}$ Meta \\ California, USA}
      
\begin{document}

\maketitle
\begin{abstract}
    Current developments in video encoding technology lead to continuously improving compression performance but at the expense of increasingly higher computational demands.
    Regarding the online video traffic increases during the last years and the concomitant need for video encoding, encoder complexity control mechanisms are required to restrict the processing time to a sufficient extent in order to find a reasonable trade-off between performance and complexity. 
    We present a complexity control mechanism in SVT-AV1 by using speed-adaptive preset switching to comply with the remaining time budget. 
    This method enables encoding with a user-defined time constraint within the complete preset range with an average precision of 8.9 \% without introducing 
    any additional latencies.
\end{abstract}
\keywords{SVT-AV1, encoder, complexity control, time control}

\section{Introduction}
Advances in computing power enable solutions to problems that weren't even solvable only a few decades ago. 
But those rapid improvements come with a cost, as keeping up with the latest technology is quite expensive for end users as well as companies. 
Thus, in order to keep on track with state-of-the-art technology to reach the best performance for compute-demanding applications, cloud computing centers provide a solution to get the current technological advancements in terms of processing power for a reasonable price~\cite{patidar2012survey} through which those services gained popularity. 
To make an economic decision concerning the trade-off between costs and computational power, it is beneficial to know the processing time of programs in advance. 
Also, being able to control the complexity to a certain degree is crucial to efficiently using the computing time in data centers.
Since video streaming is responsible for 60\% of web traffic~\cite{brandt2020} and the encoding part of video streaming has a huge computational demand, one important application field for complexity control is video encoding.

In 2018, the Alliance of Open Media (AOM) released a new coding standard, the AOMedia Video 1 (AV1)~\cite{chen2018av1}, which achieves higher bitrate savings compared to its predecessor VP9~\cite{mukherjee2013latest} but at the expense of a higher encoding time~\cite{grois2018performance}. 
Intel's open-source implementation, SVT-AV1~\cite{kossentini2020svt}, has been released one year later and reduces some of the complexity overhead.
However, the overall complexity is still high which leads to the necessity of \textit{presets}.
Those restrict the encoder optimizations in order to save computing time and enable a rough decision on the trade-off between complexity and compression quality. 
However, accurate time control cannot be achieved.

The first approach for complexity control in AV1 is introduced in~\cite{rehbein2022multi} and uses multi-objective optimization of preselected configuration parameter sets.
During encoding, a controller decides whether the current configuration is kept or changed to a faster or slower one.
With this method, complexity reductions ranging from 10\% to 40\% can be reached.
Complexity control methods in Versatile Video Coding~(VVC) restrict the coding unit (CU) partitioning process by using Support Vector Machines~\cite{wu2021svm}.
Intra complexity control is achieved by~\cite{shu2022intra} et al.\ by incorporating texture information in the CU-level partitioning process.
Also focusing on intra-coding, assigning a target time to each coding tree unit\ (CTU), and adjusting the preset accordingly~\cite{huang2022intra,huang2022precise} is another approach in VVC.
In High Efficient Video Coding~(HEVC), complexity control is achieved by adapting the depth of the CUs~\cite{correa2011complexity, correa2012adaptive} as well. 
Further improvements are achieved by e.g.\ including spatial and temporal information~\cite{correa2013coding}. 
Other methods introduce deep neural network approaches for CU partitioning~\cite{wang2021convolutional}, or CTU prediction~\cite{xu2018deepl} and control encoding complexity by pruning the weight parameters of those networks~\cite{li2020accelerate}.

This work proposes a method that enables controlling complexity in SVT-AV1 with a high accuracy within the complete preset range that is in two orders of magnitude, in contrast to other methods that often remain at a lower range.
Thereby, it bypasses the requirement of the user to set a preset by replacing this setting with a target time with only a small loss in compression performance.

In Section~\ref{sec_cc}, first, the encoding speed estimation is introduced. 
For the preset-speed relationship, a look-up table is then presented, and it is explained how it supports the preset switching decision during encoding.  
Finally, Section~\ref{sec_res} shows the target speed accuracy when encoding sequences with different resolutions and frame rates.

\section{Complexity control with quasi-continuous presets}\label{sec_cc}
A rough encoder complexity selection is already available in many encoders and is done via presets, which enable, disable, or restrict specific compression options. 
These include e.g., filters that enhance the visual quality, allowing global motion compensation, or options for block partitioning sizes.
They range from 0 to 12 for SVT-AV1, where preset 0 relates to the highest encoding quality and uses all encoder optimizations, resulting in also a high processing time. 
Vice versa, preset 12 is the option for very fast or real-time encoding solutions at a reduced compression performance. 

However, the actual processing time is highly dependent on, e.g., the video sequence and can vary in a large range.
Our work proposes a method that strives to resolve the existing rigid preset decisions and enables encoding in a quasi-continuous way within the full preset range by
\begin{enumerate}
    \item making a rough data-driven initial preset decision and
    \item switching adaptively between presets according to the current encoder speed
\end{enumerate}
in order to reach a given target time $t_{\text{target}}$. 

\subsection{Introducing Encoder Speed Feedback in SVT-AV1}
To enable controlling complexity in SVT-AV1 by speed-adaptive preset switching (SAPS), the current processing time as a measure for complexity has to be determined first. 

\subsubsection{High-level Encoder Adaptions}
In SVT-AV1, each encoding step is swapped out in one isolated process and all of them work simultaneously on the video data in a FIFO fashion. 
After the first frame is fully encoded, the measured time from the last process is transferred to the first process where an encoding speed estimate is made as well as a decision for a preset change for the upcoming frames.
To keep memory management intact, the allowed geometries for the splitting decisions are kept constant. 


\subsubsection{Estimation of the Encoding Speed}
The parallel processing within one buffer poses challenges in encoding speed calculation, i.e.\ when measuring the processing time for one frame, this value will also include the computing time of the other frames that are inside the buffer.
As a consequence, without introducing additional latencies, this value cannot be determined exactly but only estimated.

Let the current encoding speed be given as
\begin{equation}
    \label{eq::speed}
    v_{\text{enc}} = \frac{n_{\text{enc}}}{t_{\text{CPU}}},
\end{equation}
where ${t_{\text{CPU}}}$ is the accumulated time and ${n_{\text{enc}}}$ the frames that contribute to this compute time.
This value is always larger or equal to the number of fully encoded frames $n_{\text{out}}$ due to the parallel processing of buffered frames whose time also adds up to the measured value.
Finally, we count the number of incoming frames $n_{\text{in}}$ and add an estimate of the number of frames that are processed while encoding one frame, to get an approximation for the number of contributing frames:
\begin{equation}
    \label{eq::f_enc}
    n_{\text{enc}} = \frac{1}{2} \left( n_{\text{out}} + n_{\text{in}} \right).
\end{equation}
As a side note, the difference between $n_{\text{out}}$ and $n_{\text{in}}$ is equal to the processing buffer size.
Thus the number of frames in the processing loop contributing to the current time measure $n_{\text{enc}}$ is chosen as the arithmetic mean between the input frame number $n_{\text{in}}$ and completely encoded frames $n_{\text{out}}$.

Analogously to Eq.~\eqref{eq::speed}, one can define a speed budget for the remaining frames by using the total number of frames $n_{\text{total}}$:
\begin{equation}
    v_{\text{budget}} = \frac{n_{\text{total}} - n_{\text{enc}}}{t_{\text{target}} - t_{\text{CPU}}}.
    \label{eq:v_budget}
\end{equation}
This value guides the preset decision for encoding the queued frames in order to fulfill the remaining time constraint.

\subsection{Preset-Complexity Relationship}
\begin{table*}
    \begin{center}
        \begin{tabular}{c | c c c c c c c c c c c c}
            Preset & 1 & 2 & 3 & 4 & 5 & 6 & 7 & 8 & 9 & 10 & 11 & 12 \\ \hline
            Pixel rate (in kpps) & 62.6 & 119.8 & 284.3 & 564.3 & 1048 & 2610 & 4450 & 7907 & 11328 & 13664 & 17838 & 24463  \\ 
        \end{tabular}
    \end{center}        
    \caption{Look-up-table for the preset-encoding speed relations}\label{tab::look-up-table}
\end{table*}
The preset switching decision is guided by a look-up table that is created as follows:
Its initial values are created by measuring the encoding time of 13 sequences from the SJTU 4K dataset~\cite{song2013sjtu} and their downsampled versions with factors 2, 4, and 8, with QPs of 23, 27, 31, 34, presets $p \in \{1,2,\dots,12\}$, and random access (RA) configuration.
Averaging each time result $s$ over all QPs and resolutions and converting to a \textit{pixel rate}, which corresponds to a pixel-wise encoding speed with unit kilo pixel per second, leads to a look-up table consisting of a rough speed estimate for all measured presets:
\begin{equation}
    v_{\text{look-up}}(p) = \frac{1}{n_\text{SJTU}} \cdot \sum_{s=1}^{n_\text{SJTU}} \frac{W^{(s)} \cdot H^{(s)} \cdot n_{\text{total}}^{(s)}}{1000 \cdot t_{\text{total}}^{(s)}(p) } [\text{kpps}].
\end{equation}
$W$ and $H$ denote the width and height of the sequence, respectively, and the resulting values are displayed in Tab.~\ref*{tab::look-up-table}.
Since the actual speed also depends on other parameters instead of only the resolution, so far, the QP dependency is also included in form of a scaling factor:
\begin{equation}
    \gamma_{\text{QP}} = \frac{1}{1- 0.015 \cdot \left(\text{QP} - 17\right)}.
\end{equation}
The corresponding preset to a desired target speed can then be used as an initial guess for the encoder by first converting the encoding target speed $v_{\text{target}}$ into pixel rate units and then searching for the preset that is closest to the converted target speed in the look-up table.
Accordingly, the speed estimates from Eq.~\eqref{eq::speed} and Eq.~\eqref{eq:v_budget} are transformed to pixel rates as well.

Since the actual encoding speed highly depends on the sequence to be encoded, this table is updated during encoding by
\begin{multline}
    v_{\text{look-up}}(p_i) = (1 - w) \cdot v_{\text{look-up}}(p_i) \\
    + w \cdot \frac{v_{\text{enc}}}{v_{\text{look-up}}(p_{\text{avg}})} 
    \cdot v_{\text{look-up}}(p_i) \quad \forall p_i \in (1, 12)
\end{multline}
with an update weight $w$ and the average preset used during encoding $p_{\text{avg}} = \frac{1}{n_{\text{enc}}} \sum_{i=1}^{n_{\text{enc}}} p^{(i)}$.
Speed estimates for those non-discrete values are obtained by linearly interpolating the preset-speed table.

\subsection{Switching Criterion}
For each frame, $v_{\text{enc}}$ decides whether the preset needs to be increased or decreased, by comparing it with the target speed $v_{\text{target}}$.
If changing is necessary, then the required acceleration for the remaining frames when staying at the current preset $p$ is calculated as
\begin{equation}
    a(p) = \frac{v_{\text{budget}}}{v_{\text{table}}(p)}.
\end{equation}
Whenever this factor exceeds or falls below a given threshold, the preset adaption $\Delta$ is set to $+1$ or $-1$. 
For thresholds, we have selected $a(p) > 1$ for the decision to increase the speed and $a(p) < 0.9$ for a decrease, such that the encoder is systematically biased towards faster speeds which is important, e.g., in real-time encoding.

Then we decide if a higher or lower preset will result in reaching the time constraint, which is tested by calculating the acceleration factor for the potential new preset $a(p + \Delta)$.
If $a$ is high enough (for speed increment) or low enough (for speed decrement), $\Delta$ will be either kept or its absolute value will additionally be increased by one, see Algorithm.~\ref{tab::preset_changes}.
Otherwise, it will be set to $0$. 
Finally, the preset of frame $i$ is set to $p_{i} = p_{i-1} + \Delta$.


    \begin{algorithm}
        \SetAlgoLined
        \SetAlgoNoEnd
        \DontPrintSemicolon 
            \If{$a(p) > 1$}{
                \lIf{$a(p+1) > 0.5$}{
                    $\Delta \gets +1$
                }
                \lElseIf{$a(p+1) >  2$}{
                    $\Delta \gets +2$
                }
                \lElse{
                    $\Delta \gets 0$
                }
            }            
            \ElseIf{$a(p) < 0.9$}{                
                \lIf{$a(p-1) < 1.8$}{
                    $\Delta \gets -1$
                }
                \lElseIf{$a(p-1) < 0.45$}{
                    $\Delta \gets -2$
                }
                \lElse{
                    $\Delta \gets 0$ 
                }
            } 
            $p \gets p + \Delta$\\           
        \caption{Speed-adaptive preset switching (SAPS)}\label{tab::preset_changes}
    \end{algorithm}

\section{Evaluation}\label{sec_res}
The proposed method is tested on 26 sequences from class A2, A3, and A3 of the AOM Common Test Condiditions~\cite{zhao2021aom}. 
The complexity is measured as the sum of user and system time with single-core execution on an AMD 7452 processor with 2.35 GHz.

For evaluation, we introduce a desired target speed $v_{\text{target}}$ that describes the encoder speed in terms of encoded frames per second (fps).
Here, we are interested in reaching those arbitrary speed targets only, in contrast to real-time encoding methods that strive to achieve encode in recorded time.

\subsection{Validation of Speed Estimation}
The prerequisite for complexity control to work properly is correct speed feedback.
Therefore, we first analyze the speed approximation before we evaluate the actual speed control. 

\begin{table*}
    \begin{center}
        \begin{tabular}{| l || c c c c c c c c | c |}            
            \hline\diagbox{Class}{Target in fps} 
             & 16 & 8 & 4 & 2 & 1 & 0.5 & 0.25 & 0.125 & Average\\\hline\hline
            A2 (1920$\times$1080) &  $(^\star)$  & 13.7 \% & 11.5 \% & 6.8 \% & 5.0 \% & 6.3 \% & 7.6 \% & 8.8 \% & 8.5 \%\\
            A3 (1280$\times$720) & 5.6 \% & 6.4 \% & 3.7 \% & 10.0 \% & 8.9 \% & 10.1 \% & 14.1 \% & $(^\star)$ & 8.4 \%\\
            A4 (640$\times$360) & 6.3 \% & 5.6 \% & 10.4 \% & 12.4 \% & 10.5 \% & 15.8 \% & $(^\star)$ & $(^\star)$ &  10.2 \%\\\hline
            All & & & & & & & & & 8.9 \% \\ \hline
        \end{tabular}
    \end{center}
    \caption{Complexity control errors over all test classes and targets measured as the relative deviation of the target from the actual speed. Particular target speeds are in general not reachable for some classes and thus excluded from evaluation. Those are indicated with ($^\star$).}\label{tab::results}
\end{table*}

\begin{figure}
    \centering
    \includegraphics*[width=0.44\textwidth]{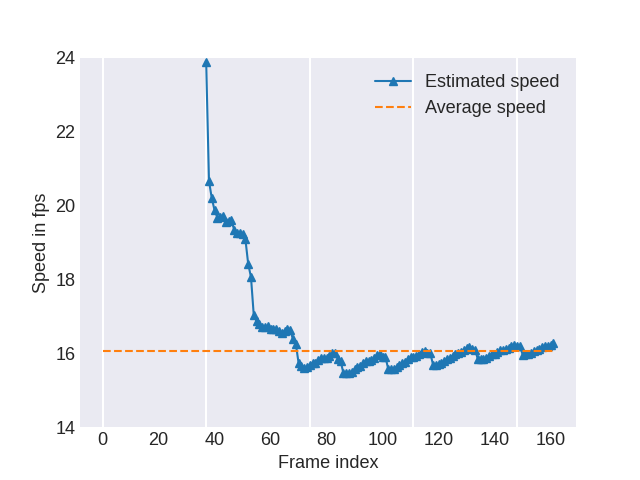}
    \caption{Validation of the speed estimation by comparing the current approximated speed with the actual average speed. The grid lines indicate the buffer sizes.}\label{img::time-error}
\end{figure}

We calculate the actual encoding speed $v_{\text{real}}$ as the number of encoded frames divided by the measured encoding time for 160 frames and compare this value with the estimated speed.
In order to use the complexity control mechanism, encoding speed should be approximately constant.
Fig.~\ref{img::time-error} shows that the method can only work properly when enough frames are processed since the time approximation is only possible after processing the first buffer and is sufficiently correct after the second. 
Buffer sizes are indicated in the figure by vertical grid lines.

\subsection{Speed Control Accuracy}
In order to ensure obtaining correct speed estimates, 300 frames are encoded for every sequence for speed control evaluation.
The encoding is performed with four constant rate factors $\text{QP} \in \left\{ 23, 27, 33, 37 \right\}$ with random access configuration, and a keyframe every 10 seconds, i.e.\ each test consists of one or two GOPs. 
Target times are set such that the encoder has a predefined encoding speed range of $v_{\text{target}} \in \left\{16, 8, 4, 2, 1, 0.5, 0.25, 0.125\right\}$ fps.
The average speed errors are calculated for each class containing $n_{\text{AOM}}$ sequences for each target speed:
\begin{equation}
    \epsilon_{\text{v} } = \frac{1}{4 \cdot n_{\text{AOM}}}\sum_{1}^{4 \cdot n_{\text{AOM}}} \frac{\left| v_{\text{real}}-v_{\text{target}} \right|}{v_{\text{target}}}
\end{equation}
The performance for each target and class is shown in Tab.~\ref{tab::results} and we obtain an average deviation over all of $8.9\%$.
Average errors of the AV1 complexity controller from~\cite{rehbein2022multi} range from 3.4\% to 11.3\% in a target complexity range from 10\% to 30\%. In comparison, the proposed method reaches relative complexities from below 1\% to 100\%.

\subsection{Complexity-Compression Performance}
\begin{figure}
    \centering
    \includegraphics*[width=0.49\textwidth]{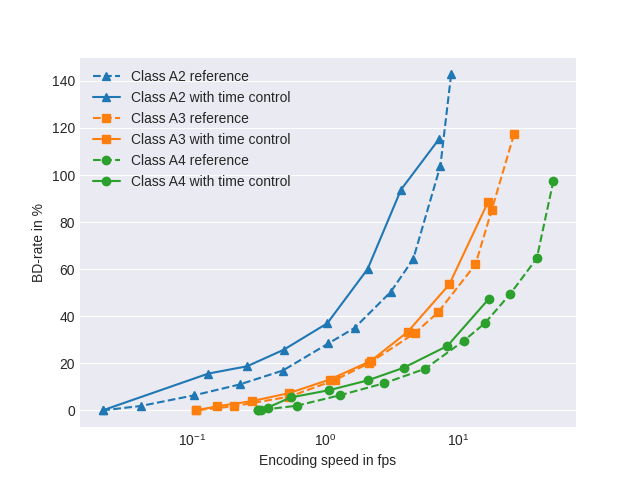}
    \caption{Trade-off between complexity and compression performance. Dashed lines denote the reference implementation over the preset range, solid lines show the proposed time control. The compression performance is calculated as the BD-rate increase compared to the anchor of preset 1, which corresponds to the slowest preset as well as the theoretical minimum reachable encoding speed in terms of speed control.}\label{img::bdbr_all}
\end{figure}

To measure the overhead in terms of compression performance loss, the Bj{\o}ntegaard Delta~\cite{bjontegaard2001calculation} bitrate (BDBR) increase compared to preset 1 is calculated for the original encoder implementation as well as the time control method.
The dependency between BDBR and encoding time over all sequences is shown in Fig.~\ref{img::bdbr_all}. 
Small resolutions (from classes A3 and A4) show a good performance regarding their trade-off since the curve of the time control method lies close to the reference curve.

\section{Conclusion}
This work proposes a complexity control approach in SVT-AV1 using the existing presets.
Over a range of two orders of magnitude we can reach target processing times with a mean error of 8.9 \%.
In future works, we will improve the time estimation for the first two buffers, which currently is the limiting factor for short video sequences, by identifying which encoding steps mostly contribute to the processing time and making speed estimations before fully encoding one frame.
Since the initial preset guess helps to reach the accurate target and to keep the rate increase low, we will also focus on introducing a more complex speed prediction model.

\bibliography{references/complexity_control}


\end{document}